\documentclass[prl,amsmath,amssymb,twocolumn,showpacs,superscriptaddress,10pt]{revtex4-1}

\pdfoutput=1

\usepackage[english]{babel}
\usepackage[utf8]{inputenc}
\usepackage[T1]{fontenc}
\usepackage{amsfonts}
\usepackage{amsthm}
\usepackage{color}
\usepackage{amsmath}
\usepackage{graphicx}
\usepackage{amssymb}
\usepackage{hyperref}
\usepackage{subcaption}
\begin{document}

\setlength{\abovedisplayskip}{5pt}
\setlength{\belowdisplayskip}{5pt}

\title{Chaotic Scattering with Localized Losses:\\ S-Matrix Zeros and Reflection Time Difference
for Systems with Broken Time Reversal Invariance}

\author{Mohammed Osman}

\affiliation{Department of Mathematics, King's College London, London WC26 2LS, United Kingdom}

\author{Yan V. Fyodorov}
\affiliation{Department of Mathematics, King's College London, London WC26 2LS, United Kingdom}
\affiliation{L.D.Landau Institute for Theoretical Physics, Semenova 1a, 142432 Chernogolovka, Russia}

\date{}

\begin{abstract}

 Motivated by recent studies of the phenomenon of Coherent Perfect Absorption, we develop the random matrix theory framework for understanding statistics of the zeros of the (subunitary) scattering matrices in the complex energy plane, as well as of the recently introduced  Reflection Time Difference (RTD). The latter plays the same role for $S-$matrix zeros as the Wigner time delay does for its poles. For systems with broken time-reversal invariance, we derive the $n$-point correlation functions of the zeros in a closed determinantal form, and study
  various asymptotics and special cases of the associated kernel.  The time-correlation function of the RTD is then evaluated and compared with numerical simulations. This allows to identify a cubic tail in the distribution of RTD, which we conjecture to be a superuniversal characteristic valid for all symmetry classes. We also discuss two methods for possible extraction of $S-$matrix zeroes from scattering data by harmonic inversion.

\end{abstract}

\maketitle

\section{Introduction}
Wave scattering in cavities with chaotic classical ray dynamics has been intensively studied over the last few decades\cite{kuhl13,grad14,diet15,hcao15}. The use of random matrix theory (RMT) has allowed for a statistical description of quantities derived directly from the $M\times M$, energy-dependent scattering matrix $S(E)$, where $M$ is the number of scattering channels; for recent reviews see \cite{fyodorov_resonance_2011, mitchell_random_2010,Schomerus2015}. In an ideal flux-conserving system $S(E)$ is unitary, but in practice unavoidable losses,  e.g. due to imperfect conductivity of the cavity walls, or leaks in connecting microwave waveguides \cite{schaefer_correlation_2003,Kuhletal2003,Hemmadi2005,Hemmadi2006a,Hemmadi2006b,microwgraphs1a,microwgraphs2,microwgraphs3}, make the experimentally observed scattering matrix subunitary. To that end, a considerable effort has been invested in generalizing the random-matrix based approaches to chaotic wave scattering in the presence of some form of absorptive loss  \cite{brouwer_voltage-probe_1997,beenbrou01,fyod03,fyodsav04,savfyodsom05,Fyo05,RFW2003,RFW2004,FF19}.
In recent years the interest in absorptive scattering has been further much stimulated by a proposal to  construct the so-called coherent perfect absorber (CPA), which can be looked at as a scattering system ( e.g. a cavity)  with a small amount of loss which completely absorbs a monochromatic wave incident at a particular frequency \cite{chong_coherent_2010}. Applications for a CPA may include optical filters and switches or logic gates for use in optical computers. Recently, a CPA in a rectangular cavity with randomly positioned scatterers and absorption due to a single antenna has been realised experimentally \cite{pichler_random_2019}, paving the way for the construction of CPAs based on disordered cavities. In another recent experiment, a CPA has been realised with a two-port microwave graph system, both with and without time-reversal symmetry \cite{chen_perfect_2020}.  In the framework of chaotic scattering, a CPA state corresponds to an eigenstate of the S matrix with zero eigenvalue at a real energy. This fact naturally motivates rising interest in a more general question of characterizing S-matrix complex zeros, which has not been systematically studied for wave chaotic systems with absorption until very recently \cite{li_random_2017,fyodorov_distribution_2017}.
This is in sharp contrast with statistics of S-matrix complex poles, known as resonances, whose
  exact density in the complex plane (and more delicate characteristics) for systems with chaotic scattering has been systematically studied in the framework of random matrix approach\cite{SokZel89,haak92,fyod96,fyodorov_statistics_1997,fyodorov_systematic_1999,somm99,jani99,
  scho00,fyod02,fyodorov_random_2003,Celardo11,FyoSav12,FyoSav15}
   with some aspects amenable to experimental verification \cite{kuhl_resonance_2008,wiersig_fractal_2008,difa12,bark13,liu14,gros14,lippolis16,DavyGenack18,DavyGenack19}.

 As is well-known, an important quantity directly related to resonance poles in a scattering system is the so-called Wigner time delay, the energy derivative of the total phase shift \cite{wigner55,smith60}, which in systems with chaotic scattering can be measured experimentally \cite{Doron1990} and whose various generalizations attracted recently much attention \cite{Rotter}. In particular, it has been suggested that complex zeroes of the scattering matrix can manifest themselves in a very analogous way via a quantity called the reflection time difference (RTD), which, in principle, may be measured experimentally \cite{fyodorov_reflection_2019}. Note that the problem of characterizing statistics of Wigner time delays and related quantities in the RMT framework (and beyond) keeps attracting considerable interest in the last 25 years \cite{lehmann_chaotic_1995,Lehmann95,FSS96,FSS97,Brouwer99,SFS01,SS03,OssFyo05,Kott05,MezSimm13,Novaes15,Cunden15,Grabsch18,Oss18,Grabsch20,GrabschTexier20}, for the reviews see \cite{fyodorov_statistics_1997} as well as the more recent \cite{Texier_td}. It is therefore natural to ask similar questions about statistics of the Reflection Time Difference in systems with chaotic scattering.

The goal of the present paper is to provide some information on fluctuations of RTD  in the framework of the RMT approach.  To this end, we mainly consider systems with broken time-reversal symmetry, where the properties
of the complex $S-$matrix zeroes can be very efficiently studied non-perturbatively (in particular, for any localized losses as well as any channel coupling) by adjusting the method suggested for $S-$matrix poles in \cite{fyodorov_systematic_1999}. This approach allows us to verify (and then exploit) that the S-matrix zeros for absorptive system form asymptotically a determinantal process in the complex plane, as long as  the effective dimension $N$ of the Hilbert space describing the cavity Hamiltonian in an appropriate energy range is considered large: $N\gg 1$.  The explicit form for the 2-point function is then used to study the RTD correlation function along the lines suggested in \cite{fyodorov_reflection_2019}, in full analogy with similar studies of the Wigner time delay \cite{FTS98}.

To begin with, let us remind that one of the most natural ways of incorporating localized losses into the RMT description is to associate them with additional (or ``hidden'') scattering channels. When those channels are numerous and weak one can model in this way spatially-uniform absorption, the idea possibly going back to \cite{brouwer_voltage-probe_1997}, cf. the discussion after equation (\ref{eq:kernel}) in the text below. For the non-perturbative localized setting the corresponding construction has been proposed in a form closest to our needs recently in  \cite{fyodorov_distribution_2017}. To this end, consider a closed cavity whose internal chaotic wave dynamics is modelled by an $N\times N$ RMT Hamiltonian $H_{0}$, coupled to $M$ scattering and $L$ absorbing channels, the latter representing the sources of localized loss. The vectors of  couplings to scattering/absorbing channels are collected in an $N\times M$ (respectively, $N\times L$) matrix $W$ ( respectively, $A$). We also define the associated $N\times N$  matrices $\Gamma_{W}=\pi WW^{\dagger}$ and $\Gamma_{A}=\pi AA^{\dagger}$, of the ranks $M$ and $L$ respectively, and further assume $M+L<N$. It is also convenient to assume that the columns of $W$ and $A$ are mutually orthogonal:
\begin{equation}\label{orthog1}
\quad\sum_{n=1}^{N}W_{na}^{*}A_{nb} = 0, \forall a=1,\ldots,M \&  b=1,\ldots,L
\end{equation}
being in addition orthogonal within each of the channel groups:
\begin{align}\label{orthog2}
    \sum_{n=1}^{N}W_{na}^{*}W_{nc} &= \gamma_{a}\delta_{ac}, \quad \sum_{n=1}^{N}A_{nb}^{*}A_{nd} = \rho_{b}\delta_{bd}.
\end{align}
 The above assumptions lead to the diagonal form of the ensemble-averaged scattering matrix, which describes only the resonant scattering associated with the creation of long-lived intermediate states. The condition (\ref{orthog2}) can be easily lifted (see e.g. Appendix A in \cite{FF19} for a recent discussion and further references).

The construction of the energy dependent flux-conserving  $(M+L)\times(M+L)$ scattering matrix $\mathcal{S}$
is done following the standard procedure frequently referred to as the ``Heidelberg approach" and going back to the seminal work \cite{VWZ85}, see also \cite{fyodorov_statistics_1997}. Adapting it to the present situation one gets
 the following block form, cf. \cite{fyodorov_distribution_2017}:
\begin{align}\label{eq:total_smat}
    \mathcal{S} &= \begin{pmatrix}
    1_{M}-2\pi iW^{\dagger}D^{-1}W&-2\pi iW^{\dagger}D^{-1}A\\
    -2\pi iA^{\dagger}D^{-1}W&1_{L}-2\pi iA^{\dagger}D^{-1}A
    \end{pmatrix},
\end{align}
where we denoted
\begin{equation}
  D(E) = E1_{N}-H_{0}+i(\Gamma_{W}+\Gamma_{A})\,.
\end{equation}

The upper left block $S(E):= 1_{M}-2\pi iW^{\dagger}D^{-1}W$ describes the scattering between $M$ ``observable'' channels and has the alternative representation:
\begin{align}\label{eq:smat}
    S(E) &= \frac{1_{N}-iK_{A}}{1_{N}+iK_{A}},\quad
    K_{A} &= \pi W^{\dagger}\frac{1}{E-H_{0}+i\Gamma_{A}}W.
\end{align}
Note that due to the presence of hidden/absorbing channels encapsulated via $\Gamma_{A}\ne 0$ the matrix
$S(E)$ is subunitary reflecting the loss of flux injected through the observable channels which escapes via
the hidden channels, and as such is treated as irretrievably absorbed. A similar representation holds for the lower right block $S'(E)$ after the replacement $W\leftrightarrow A$ everywhere.

 Since $S(E)$ is subunitary, positions of its zeros in the complex energy plane are no longer conjugates of the corresponding poles, and thus in principle can be located in both half-planes. From \ref{eq:smat} one can easily deduce that the determinant of $S(E)$ has the following form:
\begin{align}
    \det S(E) &= \frac{\det[E1_{N}-H_{0}+i(\Gamma_{A}-\Gamma_{W})]}{\det[E1_{N}-H_{0}+i(\Gamma_{W}+\Gamma_{A})]},
\end{align}
from which it is clear that the zeros $z_{n}$ of $S(E)$ in the complex energy plane are the complex eigenvalues of the non-Hermitian matrix $H_{0}+i(\Gamma_{W}-\Gamma_{A})$. Writing a similar expression for $\det S'(E)$ and taking their ratio, we arrive at the following complex number:
\begin{align}
    \frac{\det S(E)}{\det S'(E)} &= \frac{\det[E1_{N}-H_{0}+i(\Gamma_{A}-\Gamma_{W})]}{\det[E1_{N}-H_{0}+i(\Gamma_{W}-\Gamma_{A})]}\\
    &= e^{2i\phi(E)}.
\end{align}
which is obviously unimodular for real values of the energy $E$.

Now we follow the proposal of \cite{fyodorov_reflection_2019} and define the reflection time difference (RTD) as the energy derivative of the phase $\phi(E)$:
\begin{align}
    \delta\mathcal{T}(E) &:= -i\frac{\partial}{\partial E}\log\frac{\det S(E)}{\det S'(E)}\\
    &= \sum_{n=1}^{N}\frac{2\Im z_{n}}{(E-\Re z_{n})^{2}+(\Im z_{n})^{2}} \label{eq:rtd}.
\end{align}
Such a definition of RTD is inspired by the Wigner time delay, which has the same form as \ref{eq:rtd} but with the zeros $z_{n}$ replaced by the complex $S-$ matrix poles located in the lower half-plane of complex energies.
Those are simply complex eigenvalues of another non-Hermitain matrix ${\cal H}_{eff}:=H_{0}-i(\Gamma_{W}+\Gamma_{A})$ and will be denoted
$\mathcal{E}_{n}=E_{n}-i\Gamma_{n}/2$, with condition $\Im{\mathcal E}_n=-\Gamma_{n}<0$ due to non-negativity of $\Gamma_{W}+\Gamma_{A}$. The main difference between the RTD $\delta\mathcal{T}(E)$ and the Wigner time delay is that the former can be negative whereas the Wigner time delay is always positive in the present model in view of
$\Gamma_{n}>0$. The name reflection time difference comes from the fact that the phase of $\det S(E)$ gives the delay (averaged over the scattering channels) in the propagation of a nearly monochromatic wave due to scattering, relative to a perfectly reflecting cavity. Hence the RTD is the difference in this delay between the first $M$ channels, deemed observable, and the last $L$ channels, deemed absorbing.
Let us however stress, that equivalently in any flux-conserving two-terminal system one can always
simply subdivide channels into two groups; most naturally, in a  two-terminal scattering setup, via the left/right division. Then  $S(E)$ and $S'(E)$ in such a system simply describe reflection blocks of the total $S-$matrix. Note that RMT-based statistics of entries and eigenvalues of the reflection blocks is interesting in itself, and being not unrelated to Wigner time delays have been studied from various viewpoints, e.g. in  \cite{fyod03,SS03,Grabsch20,Kanzi12,direct}.


\section{Statistics of S matrix zeros for systems with broken time reversal invariance}

 We have thus seen that the zeros $z_{n}$ of $S(E)$ in the present approach are nothing else but the complex eigenvalues of the non-Hermitian $N\times N$ random matrix $H_{0}+i(\Gamma_{W}-\Gamma_{A})$. As the zeroes are the main constituents of the RTD (\ref{eq:rtd})
 we briefly analyze statistics of those zeroes in the complex
plane for wave-chaotic systems with broken time-reversal invariance, when the matrix $H_{0}$ is taken from the Gaussian Unitary Ensemble (GUE). Note that systems of such type can be studied experimentally, see e.g. \cite{LawSir2018} and references therein.

 Since we have assumed the orthogonality both inside and between the two groups of channels, see (\ref{orthog1})-(\ref{orthog2}), exploiting unitary invariance of the GUE part we can easily check that the matrix $\Gamma:=\Gamma_{W}-\Gamma_{A}$ of rank $M+L<N$ can be chosen to be diagonal: $\Gamma=\text{diag}(\gamma_{1},...,\gamma_{M},-\rho_{1},...,-\rho_{L},0,...,0)$. In the lossless case $L=0$, $\Gamma$ is necessarily positive and all $n$-point correlation functions have been derived in \cite{fyodorov_systematic_1999} in the limit $N\gg \max(M,n)$. We show in the Appendix which modification are necessary to adapt the method \cite{fyodorov_systematic_1999} to the lossy case $L>0$. The end result is the following determinantal form for the asymptotic $n$-point correlation functions for the  eigenvalues $z_n$ in the whole complex energy plane:

\begin{align}\label{clusterfunctions}
    \lim_{N\to\infty}\frac{1}{N^{2n}}R_{n}\left(x+\frac{z_{1}}{N\pi\nu(x)},...,x+\frac{z_{n}}{N\pi\nu(x)}\right)
\end{align}
\[
=\det[K(z_{i},z^{*}_{j})]_{1\leq i,j\leq n},
\]
where $x\in(-2,2)$, $\nu(x)=\frac{1}{2\pi}\sqrt{4-x^{2}}$ is the semicircular density of real eigenvalues of
the GUE matrix $H_0$, and the kernel is given explicitly by:
\begin{align}\label{eq:kernel}
    K(z,w^{*}) = \sqrt{F(z)F(w^{*})}
\end{align}
\[
\times \int_{-1}^{1}du\,e^{i(z-w^{*})u}\,\prod_{a=1}^{M}(g_{a}+u)\prod_{b=1}^{L}(h_{b}-u),
\]
where for $z=\Re z+i\Im z$ we have
\begin{align}
    F(z) &= \int_{-\infty}^{\infty}\frac{dk}{2\pi}\frac{e^{-2ik\Im z}}{\prod_{a=1}^{M}(g_{a}-ik)\prod_{b=1}^{L}(h_{b}+ik)},\\
    g_{a} &= \frac{1}{2\pi\nu(x)}\left(\gamma_{a}+\frac{1}{\gamma_{a}}\right),\quad h_{b}=\frac{1}{2\pi\nu(x)}\left(\rho_{b}+\frac{1}{\rho_{b}}\right)\,.
\end{align}

Let us first check the simplest limit of very many equivalent weakly coupled absorbing channels, namely $L\to \infty, \rho_b \to 0$ in such a way that the product $L\rho_b$ remains a finite constant.  Following
\cite{brouwer_voltage-probe_1997} one expects that the absorption becomes spatially uniform across the sample, and that  all zeroes $z_n$ will be uniformly shifted downwards in the complex plane by the same amount. Indeed, introducing the notation $L\rho_b=\epsilon/(\pi\nu(x)), \, \forall b=1,\ldots, L$ and assuming it to remain constant as $L\to \infty$, we see that we can replace
$h_{b}\approx \frac{L}{2\epsilon}$, and easily verify that the kernel in (\ref{eq:kernel}) reduces to the $L=0$ case but with the shift $\Im z\to \Im z+\epsilon$, in full correspondence with the uniform absorption picture.

A less trivial, representative case to be considered next is that of equivalent scattering channels $g_{a}=g,\, \forall a=1,\ldots,M$, as well of equivalent absorbing channels $h_{b}=g_{0},\,\forall b=1,\ldots,L$, but both couplings not considered to be weak. The interesting limiting case arises if we again assume the channels are abundant, so that both  $M\to\infty$ and $L\to \infty$, but in such a way that the ratio
  $\frac{L}{M}=p$ with $0\leq p\leq1$ remaining constant (the case $p>1$ follows by replacing $p\to\frac{1}{p}$ and  $M\leftrightarrow L$ and $g\leftrightarrow h$). In such a limit the integrals over variables $u$ and $k_{1,2}$ can be readily evaluated by the Laplace (saddle-point) method. The calculation can be performed for general $p,g,g_{0}$ but the resulting expressions are relatively cumbersome; here we present explicit formulas only for the special case $p=1$ and $g=g_{0}$ when they are more elegant. The ensuing asymptotic mean density of complex zeroes is supported inside the domain
  \begin{equation}\label{eq:boundary1}
  \left\{(x,y):-2\leq x\leq 2, -\frac{M}{g^{2}-1}\leq y\leq \frac{M}{g^{2}-1}\right\}\,.
   \end{equation}
   Note that since $g$ depends on $x$ as $g\sim 1/\nu(x)$, the width of the support along the imaginary axis decreases monotonically to zero at $x\to \pm2$. Inside the support, the density is constant near the real axis and decays as $y^{-2}$ when $y=O(M)$: defining $\widetilde{\rho}(x,y) := \frac{\rho(x,y)}{\nu(x)}$ we then have
\begin{align}
    \widetilde{\rho}(x,y) &= \begin{cases}
    \frac{g^{2}}{2\pi M}&\quad \mbox{if}\,\, y=O(1)\\
    \frac{g^{2}}{4\pi M\widetilde{y}^{2}}\left(1+\frac{1}{\sqrt{1+4g^{2}\widetilde{y}^{2}}}\right)&\quad  \mbox{if}\,\,y=M\widetilde{y}
    \end{cases}\label{eq:density}.
\end{align}

For the general parameters $p,g,g_0$ the support of the density in the complex plane is given for  $-2\leq x\leq 2$ by
\begin{align}
    -\frac{M}{2}\left(\frac{p}{g_{0}-1}-\frac{1}{g+1}\right)\leq y\leq\frac{M}{2}\left(\frac{1}{g-1}-\frac{p}{g_{0}+1}\right).\label{eq:boundary}
\end{align}

\begin{figure}[!ht]
    \centering
    \begin{subfigure}{\linewidth}
        \includegraphics[width=\linewidth,keepaspectratio=true]{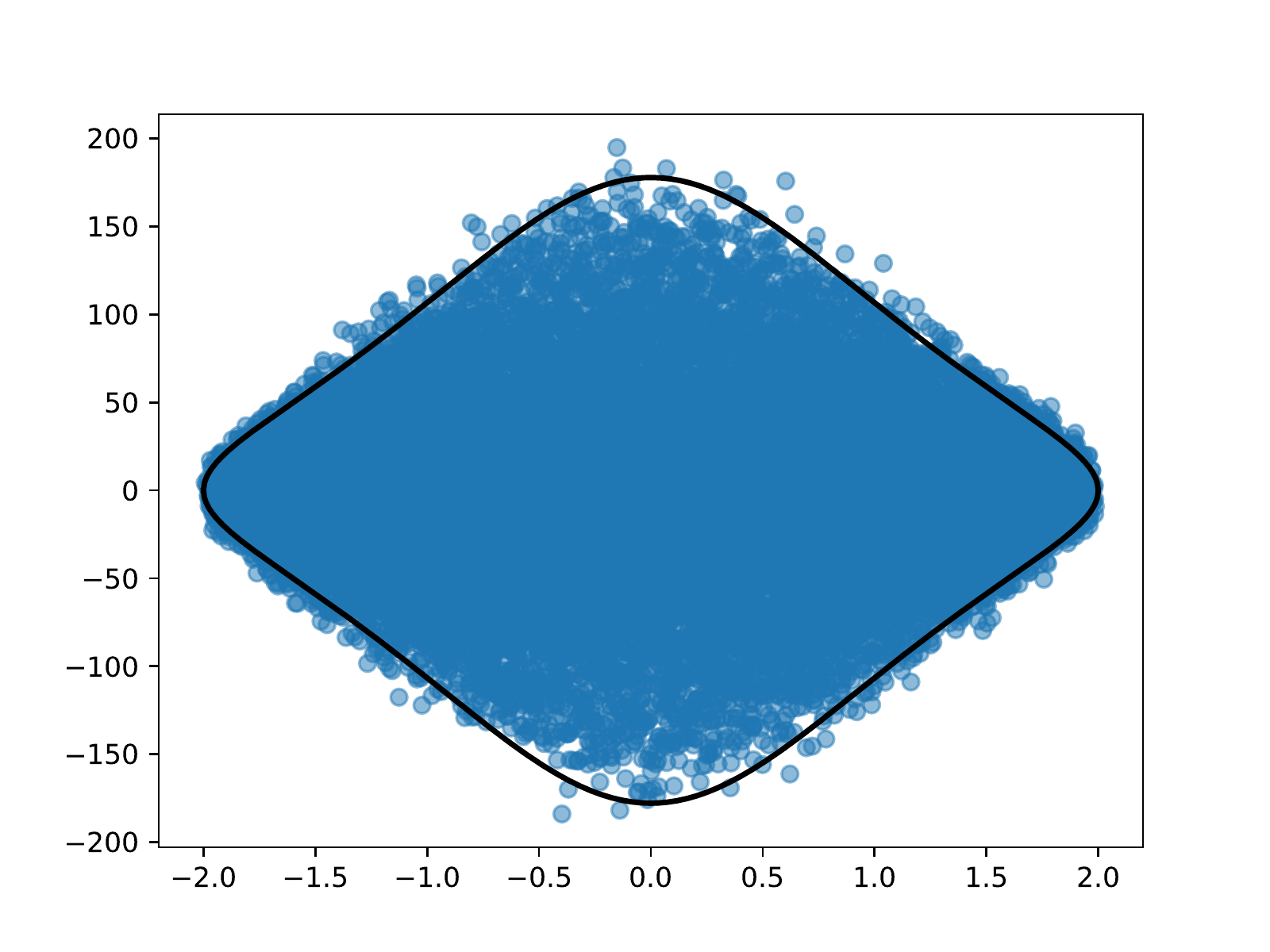}
        \caption{\small $M=L=100$ and $g=g_{0}=1.25$}
        \label{fig:support1}
    \end{subfigure}\\
    \begin{subfigure}{\linewidth}
        \includegraphics[width=\linewidth,keepaspectratio=true]{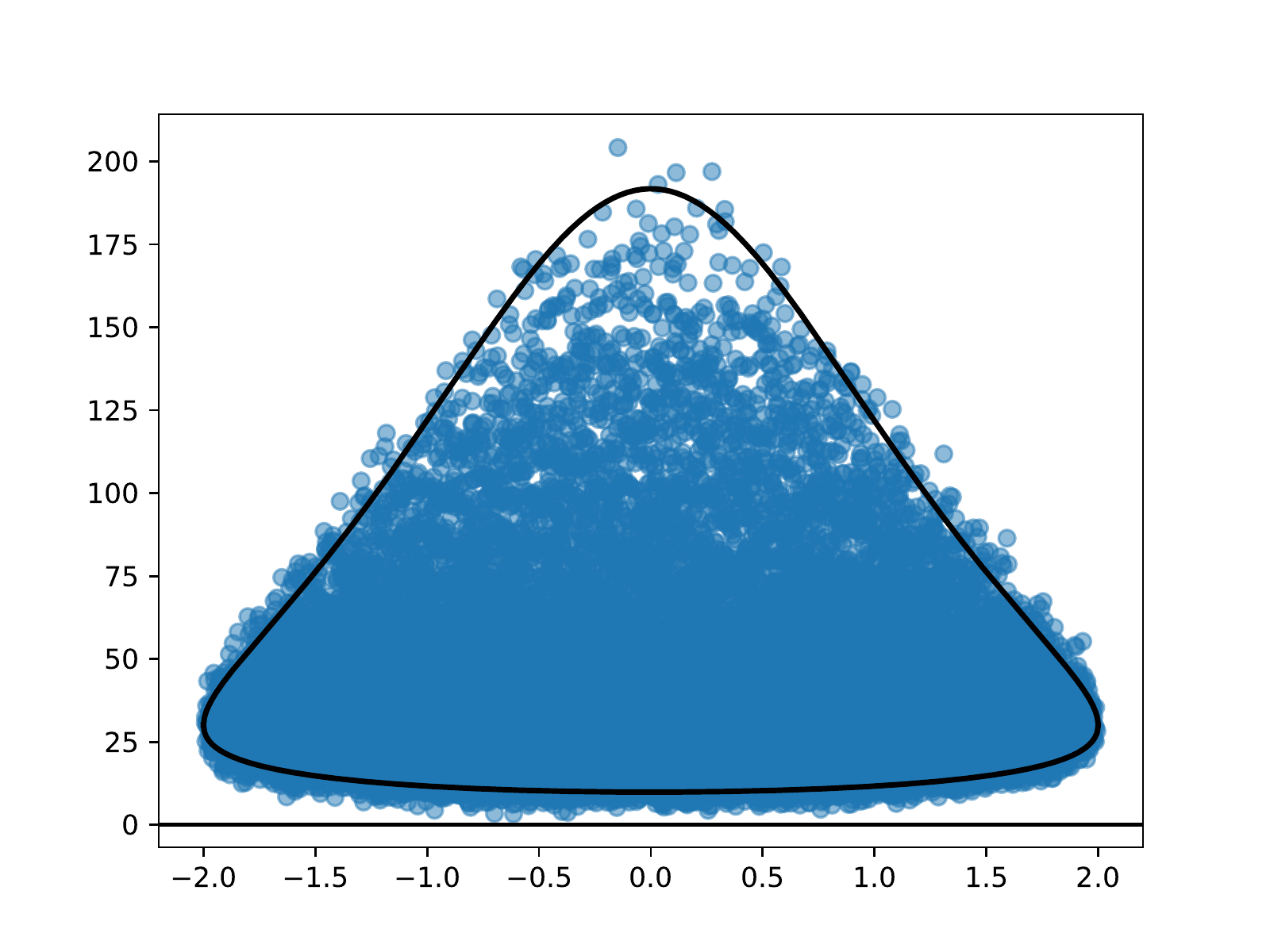}
        \caption{\small $M=L=100,\,g=1.25$ and $g_{0}=5.05$}
        \label{fig:support2}
    \end{subfigure}
    \caption{\small Eigenvalues of five $4000\times4000$ matrices with the solid line indicating equations \ref{eq:boundary1} (top) and \ref{eq:boundary} (bottom)}
    \label{fig:support}
\end{figure}

Figures \ref{fig:support1} and \ref{fig:support2} show the eigenvalues of five $4000\times4000$ matrices with $M=L=100$ and $g=g_{0}=1.25$ and $g=1.25,\,g_{0}=5.05$ respectively.

In particular, we see that if $p=0$ then the zeros are all in the upper half-plane, separated from the real axis by the gap $\frac{M}{2(g+1)}$. Remembering that for $p=0$ zeroes are mirror images of poles, this result is simply $L=0$ case of \cite{fyodorov_systematic_1999}. The gap in the poles distribution is the well-known feature first derived in \cite{haak92} and \cite{lehmann_chaotic_1995}
in the related, but slightly different limit $M=O(N)$. It has profound consequences for underlying dynamics
and also has a semiclassic significance, being related to the classic escape time\cite{GaspardRice89a,GaspardRice89b}.
Moreover, one can see that all the S-matrix zeros will still  be in the same half-plane as long as $|pg-g_{0}|>1+p$ (upper half-plane for $g_{0}-pg>1+p$ and lower for $pg-g_{0}>1+p$). An illustration of such a situation is given in the figure \ref{fig:support2}.

When $|z_{1}-z_{2}|=O(1/N)$, the kernel can be reduced after some algebraic manipulations to the Ginibre-like form:
\begin{align}
    |K(z_{1},z_{2}^{*})| &= \widetilde{\rho}(z)e^{-\frac{\pi}{2}\widetilde{\rho}(z)|z_{1}-z_{2}|^{2}},
\end{align}
where $z=\frac{z_{1}+z_{2}}{2}$ and $\widetilde{\rho}(z)$ is from \ref{eq:density}. Such a kernel was found in the $L=0$ case when $M\to\infty$ \cite{fyodorov_systematic_1999} and is conjectured to be the universal form of the kernel for strongly non-Hermitian matrices \cite{fyodorov_random_2003}.

We also record for completeness the general expression for the density of zeros, valid for any $M,L$:

\begin{align}
   & \widetilde{\rho}(x,y) =\frac{1}{(g+g_{0})^{M+L-1}}\,\int_{-1}^{1}due^{-2uy}(g+u)^{M}(g_{0}-u)^{L}  & \label{eq:density_general}\nonumber\\ \times   &\left\{\theta(-y)e^{2g_{0}y}\sum_{n=1}^{L}a_{n}(L,M)\frac{[-2(g+g_{0})y]^{n-1}}{\Gamma(n)}\right.&\nonumber\\
    &\quad\left.+\theta(y)e^{-2gy}\sum_{n=1}^{M}a_{n}(M,L)\frac{[2(g+g_{0})y]^{n-1}}{\Gamma(n)}\right\},&
\end{align}
where
\begin{equation}\label{an}
a_{n}(M, L) = \frac{\Gamma(M+L-n)}{\Gamma(M-n+1)\Gamma(L)}.
\end{equation}

\begin{figure}[!ht]
    \centering
    \begin{subfigure}{\linewidth}
        \includegraphics[width=\linewidth,keepaspectratio=true]{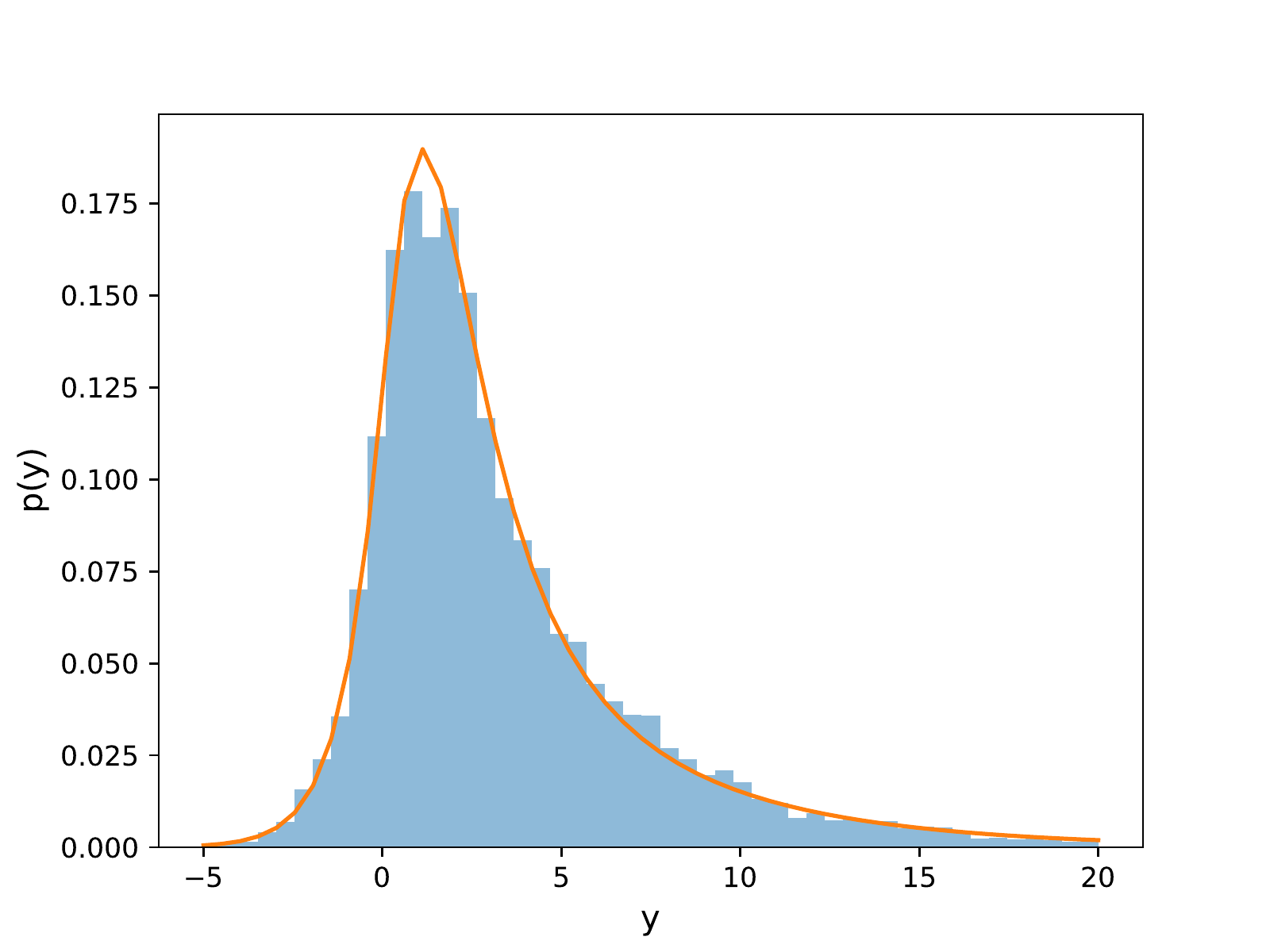}
        \caption{\small $N=100,\,M=4,\,L=2$}
        \label{fig:density_a}
    \end{subfigure}\\
    \begin{subfigure}{\linewidth}
        \includegraphics[width=\linewidth,keepaspectratio=true]{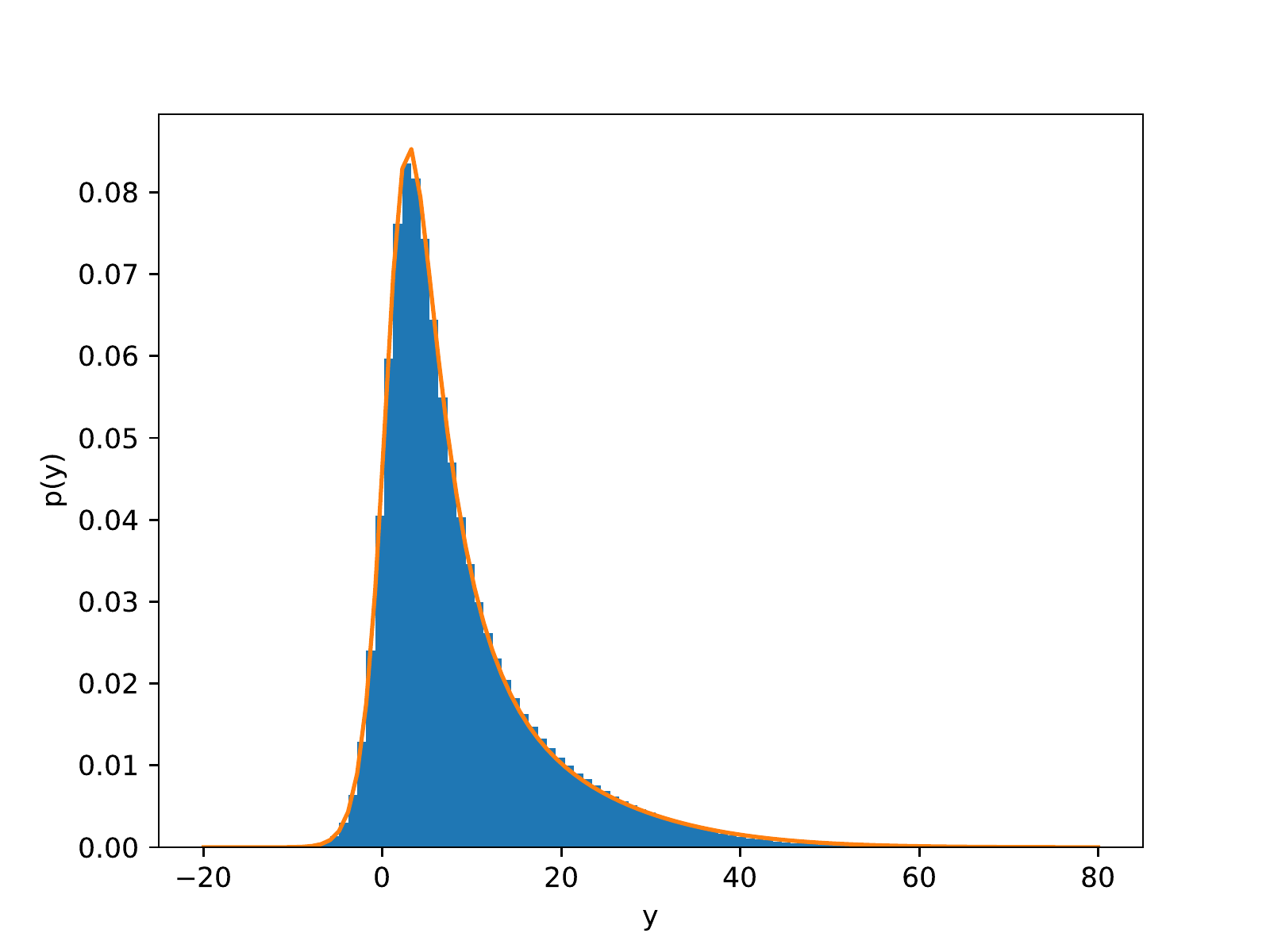}
        \caption{\small $N=500,\,M=10,\,L=5$}
        \label{fig:density_b}
    \end{subfigure}
    \caption{\small Density of the imaginary parts $\Im z_{n}$ for $\Re z_{n}=0$ compared against \ref{eq:density_general} (solid line).}
    \label{fig:density}
\end{figure}

The above mean density in $y$ for $x=0$ is compared with RMT simulations in figures \ref{fig:density_a} and \ref{fig:density_b}.

\section{Statistics of Reflection Time Difference for systems with broken time reversal invariance}

In this section we convert the information about $S-$matrix zeroes into information about
the two-point connected correlation function of the Reflection Time difference $\delta\mathcal{T}$
defined via (\ref{eq:rtd}). In doing this we largely follow the method proposed in \cite{FTS98} for
the Wigner time delays. We start with recalling that the main microscopic energy scale characterizing (real) eigenvalues of the Hermitian RMT cavity Hamiltonian $H_{0}$ is the associated mean level spacing $\Delta=[N\nu(E)]^{-1}$, and introduce the appropriately rescaled RTD via $\widetilde{\delta\mathcal{T}}=\frac{\Delta}{2\pi}\delta\mathcal{T}$.
From the technical point of view
it is more natural to consider the associated Fourier transform of the two-point correlation function in question defined as
\begin{align}\label{corrRTD1}
   C_{M,L}(t) &:=   \frac{1}{2\pi}\int d\omega\,e^{i\omega t}\langle\widetilde{\delta\mathcal{T}}\left(E+\frac{\Delta}{2\pi}\omega\right)
   \widetilde{\delta\mathcal{T}}\left(E-\frac{\Delta}{2\pi}\omega\right)\rangle_{c}.
 \end{align}
 Defining the so-called empirical density of $S-$matrix zeroes $z_n$ in the complex plane $z=\Im z+i\Re z$ as
 \begin{equation}\label{empden}
 \rho(z)=\frac{1}{N}\sum_{n=1}^N\delta^{(2)}(z-z_n),
 \end{equation}
 where $\delta^{(2)}(z-z_n):=
 \delta(\Re z-\Re z_n)\delta(\Im z-\Im z_n)$, with $\delta(x)$ being the standard Dirac delta-function, we first rewrite (\ref{eq:rtd}) as
 \begin{equation}
 \widetilde{ \delta\mathcal{T}}(E)=\frac{\Delta\, N}{2\pi}\int \frac{2\Im z}{(E-\Re z)^{2}+(\Im z)^{2}}\,\rho(z) \,\,d\Re z\,d\Im z,
 \end{equation}
 and then substitute this in (\ref{corrRTD1}) and perform the ensemble averaging. In doing this we
 exploit the relation between covariance functions of empirical densities and the two-point function featuring in
 (\ref{clusterfunctions}) for $n=2$, hence the associated kernel (\ref{eq:kernel}):
 \begin{equation}
 \left\langle\rho(z_1)\rho(z_2)\right\rangle_c=\left\langle\rho(z_1)\right\rangle\delta(z_1-z_2)-\mathcal{Y}_{2}(z_{1},z_{2}),
 \end{equation}
 where $\mathcal{Y}_{2}(z_{1},z_{2})=|K(z_{1},z_{2}^{*})|^{2}$ is the associated two-point ``cluster" function.
   After rescaling and straightforward manipulations this brings (\ref{corrRTD1}) to the form
 \begin{equation}\label{corrRTD2}
 C_{M,L}(t>0)= A(t)-B(t), \quad A(t)=\frac{1}{2}\int dye^{-2\lvert y\rvert t}\widetilde{\rho}(E,y),
 \end{equation}
  and
 \begin{align}
   B(t)= \frac{1}{2\pi}\int d\omega dy_{1}dy_{2}\,e^{-it\omega-(\lvert y_{1}\rvert+\lvert y_{2}\rvert)t}\label{eq:rtd_corr}
 \end{align}
 \[
\times  \mathcal{Y}_{2}(E,\omega,y_{1},y_{2})\,\text{sign}(y_{1}y_{2}),
 \]
   Finally, using the expression \ref{eq:kernel} for the kernel, we arrive at the following formulas:

\begin{eqnarray}
    & A(t) & =\frac{1}{4}\int_{-1}^{1}du\frac{(g+u)^{M}(g_{0}-u)^{L}}{(g+g_{0})^{M+L}}\nonumber \\
   & &\quad\quad\times\left[\sum_{n=1}^{M}a_{n}(M,L)\left(\frac{g+g_{0}}{g+t+u}\right)^{n}\right. \nonumber\\
   & &\quad\quad\left.+
    \sum_{n=1}^{L}a_{n}(L,M)\left(\frac{g+g_{0}}{g_{0}+t-u}\right)^{n}\right],
    \end{eqnarray}
   \begin{eqnarray}
    && B(t) = \frac{\theta(2-t)}{4}\int_{-1}^{1-t}du\frac{(g+u)^{M}(g_{0}-u)^{L}}{(g+g_{0})^{2(M+L)}} (g+t+u)^{M}
    \nonumber\\ &&
    \qquad\times (g_{0}-t-u)^{L}\left[\sum_{n=1}^{M}a_{n}(M,L)\left(\frac{g+g_{0}}{g+t+u}\right)^{n} \right. \nonumber \\ && \left. \quad\quad -
    \sum_{n=1}^{L}a_{n}(L,M)\left(\frac{g+g_{0}}{g_{0}-u}\right)^{n}\right]^{2} .
\end{eqnarray}
where $a_{n}(L,M)$ has been defined in (\ref{an}).

A few remarks are due. First, the limit of no absorption is equivalent to sending $g_{0}\to\infty$. The dominant
contribution then obviously comes from the $n=M$ term in the first sum,  in both $A(t)$ and $B(t)$. The resulting expression reproduces the well-known correlation function of the Wigner time delay for systems with broken time-reversal invariance: \cite{FSS96,fyodorov_statistics_1997}:
\begin{align}
    C_{M,0}(t) &= \frac{1}{4}\int_{\text{max}(1-t,-1)}^{1}\left(\frac{g+u}{g+t+u}\right)^{M}du,
\end{align}

Note that the above correlation function of Wigner time delays decays as $t^{-M}$ for $t\to \infty$, whereas the corresponding correlation of RTD for any finite absorption $g_0<\infty$ decays in the same limit as $t^{-1}$ for any fixed value $M>0,L>0$. This implies that the variance of the RTD which in view of (\ref{corrRTD1}) can be found as
\[
\left\langle\left[\widetilde{\delta\mathcal{T}}\left(E\right)\right]^2\right\rangle_{c}\propto \int_{0}^{\infty}C_{M,L}(t)dt,
\]
  logarithmically diverges for any finite number of absorbing channels. This feature is strikingly different from the Wigner time delay, whose variance is infinite only for a single-channel scattering, being finite for any $M>1$. The logarithmic divergence of the variance suggests that the distribution of RTD must have
  the large-tail behaviour  $P(\widetilde{\delta\mathcal{T}})\sim \widetilde{\delta\mathcal{T}}^{-3}$.
  Although we can not prove this conjecture rigorously in generality, it can be strongly supported
  by a perturbative argument (sketched in the Appendix) valid in the regime of small absorption in a weakly open system with equivalent channels, $\gamma_a=\gamma\ll 1, \rho_a=\rho\ll 1$,  when imaginary parts of the $S-$matrix zeros are much smaller than their separation: $\langle|\Im z_{n}|\rangle\ll\Delta$.  Adapting the argument along the lines used in \cite{fyodorov_statistics_1997} for Wigner time delays, and further assuming 
  $\rho\gtrsim \gamma$, the probability density for RTD  can be shown to have an algebraic decay with universal exponents,
\begin{align}\label{unitail}
    P_{\beta}(\widetilde{\delta\mathcal{T}}) &\sim \begin{cases}
    \lvert\widetilde{\delta\mathcal{T}}\rvert^{-3/2}, &1\ll \lvert\widetilde{\delta\mathcal{T}}\rvert\ll \gamma^{-1}\\
    \lvert\widetilde{\delta\mathcal{T}}\rvert^{-3}, &\lvert\widetilde{\delta\mathcal{T}}\rvert\gg \gamma^{-1}
    \end{cases},
\end{align}
and similarly for negative $\widetilde{\delta\mathcal{T}}$ with $\gamma$ and  $\rho$ exchanging their roles. This result is ``superuniversal'', that is holding not only for systems with broken symmetry, but for all standard symmetry classes of $H_{0}$ described by values of the Dyson index  $\beta=1,2,4$ and for all $M>0,L>0$. We indeed see that the infinite variance is due to the cubic asymptotic decay in the probability density. Note that in the case of Wigner time delay $\tau$ the asymptotic tail of the probability density is rather $\tau^{-\beta M/2-2}$ making the variance finite for $M>2/\beta$. Anticipating that for $\rho=0$ RTD should be indistinguishable from the Wigner time delay, one may consider the parameter range $\rho\ll \gamma$ and find that in that case the decay
 $\widetilde{\delta\mathcal{T}}^{-\beta M/2-2}$ also happens for RTD in the intermediate asymptotic range
 $ 1/\gamma\ll \widetilde{\delta\mathcal{T}}\ll 1/\rho$, whereas the cubic tail takes over only as
  $\widetilde{\delta\mathcal{T}}\gg 1/\rho$, reconciling the two types of behaviour. 

One also can be interested in the short-time behaviour of the RTD correlation function.
Considering for simplicity the simplest case $M=L=1$, we obtain in this limit:

\begin{align}
    C_{1,1}(t\ll 1) &= \frac{1}{2}\left[1-\frac{4+3(g-g_{0})^{2}}{3(g+g_{0})^{2}}\right]\nonumber
    \\ &+\frac{(g-g_{0})^{2}-4(g+g_{0}-1)}{4(g+g_{0})^{2}}t+O(t^{2}).
\end{align}

The first order term vanishes when $g^{\pm}_{0}=g+2\pm2\sqrt{2g}$ and vice versa. For $1<g<3+2\sqrt{2}$, the second solution $g^{-}_{0}=g+2-2\sqrt{2g}<1$ is not valid since $g,g_{0}>1$ by definition. Thus, the correlator switches from an increasing to a decreasing function of time as $g_{0}$ passes through $g_{0}^{+}$. Reverting to the frequency domain, we find that in this case correlator decays as $\omega^{-4}$, whereas the correlator of the Wigner time delay always decays as $\omega^{-2}$ since $\dot{C}_{M,0}(0)=1/4$.

Another curious observation is that due to appearance of the factor $\text{sign}(y_{1}y_{2})$ in the integrand of (\ref{eq:rtd_corr1}), the term $B(t)$ identically vanishes when $\mathcal{Y}_{2}$ is even in $y_{1}$ and $y_{2}$ separately, in which case $C(t)$ only depends on the mean global density of complex eigenvalues rather than
on the two-point correlation function. For equivalent channels, a necessary condition for this to happen is $M=L$ and $g=g_{0}$.

To compare with numerical simulations, we find that instead of directly computing \ref{corrRTD1}, it is more practical to consider instead the Fourier transform of the RTD weighted by a Gaussian:

\begin{align}\label{eq:rtd_corr1}
    F(t,W) &:= \int\frac{dE}{\sqrt{2\pi W}}\widetilde{\delta\mathcal{T}}(E)e^{iEt-\frac{E^{2}}{2W}}\\
    &=\sqrt{\frac{\pi}{2W}}e^{-\frac{Wt^{2}}{2}}\sum_{n=1}^{N}\text{sign}(\Im z_{n})\nonumber\\
    &\times\left[\text{erfcx}\left(\frac{|\Im z_{n}|+Wt+i\Re z_{n}}{\sqrt{2W}}\right)\right.\nonumber\\
    &+\left.\text{erfcx}\left(\frac{|\Im z_{n}|-Wt-i\Re z_{n}}{\sqrt{2W}}\right)\right],
\end{align}

where $\text{erfcx}(z)=e^{z^{2}}\text{erfc}(z)$ and $\text{erfc}(z)=\frac{2}{\sqrt{\pi}}\int_{z}^{\infty}e^{-x^{2}}dx$ is the complementary error function. If we take $\frac{1}{N}\ll\sqrt{W}\ll1$, then we find the approximate relation:

\begin{align}\label{eq:rtd_corr2}
    \frac{\Delta}{4\pi}\sqrt{\frac{W}{\pi}}\langle|F\left(\frac{\pi t}{\Delta};W\right)|^{2}\rangle_{c} &\approx C_{M,L}(t).
\end{align}

\begin{figure}[!ht]
    \centering
    \begin{subfigure}[t]{\linewidth}
        \includegraphics[width=\linewidth,keepaspectratio=true]{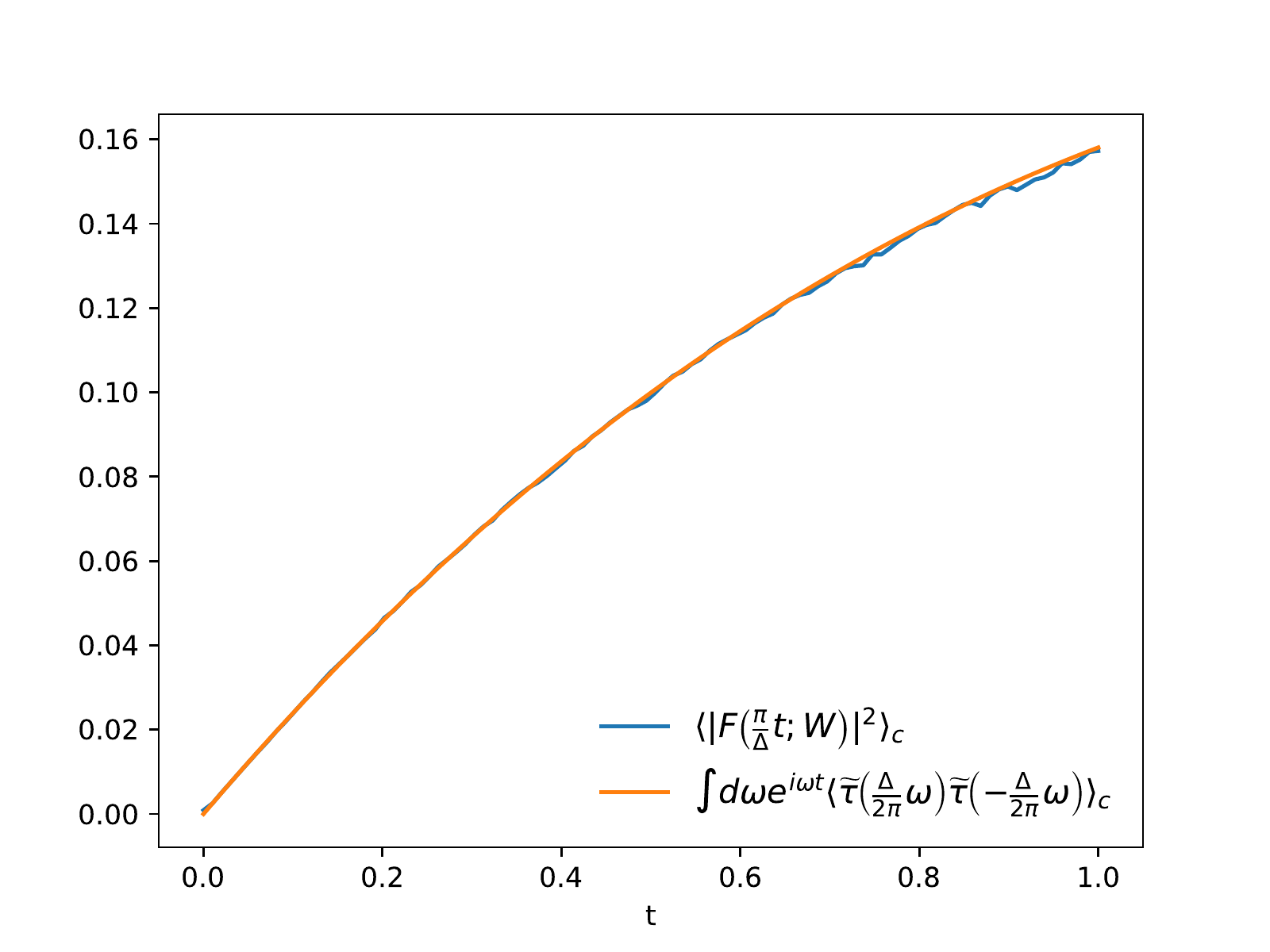}
        \caption{\small $M=1,L=0,g=1.25$}
        \label{fig:rtd_corr1}
    \end{subfigure}\\
    \begin{subfigure}[t]{\linewidth}
        \includegraphics[width=\linewidth,keepaspectratio=true]{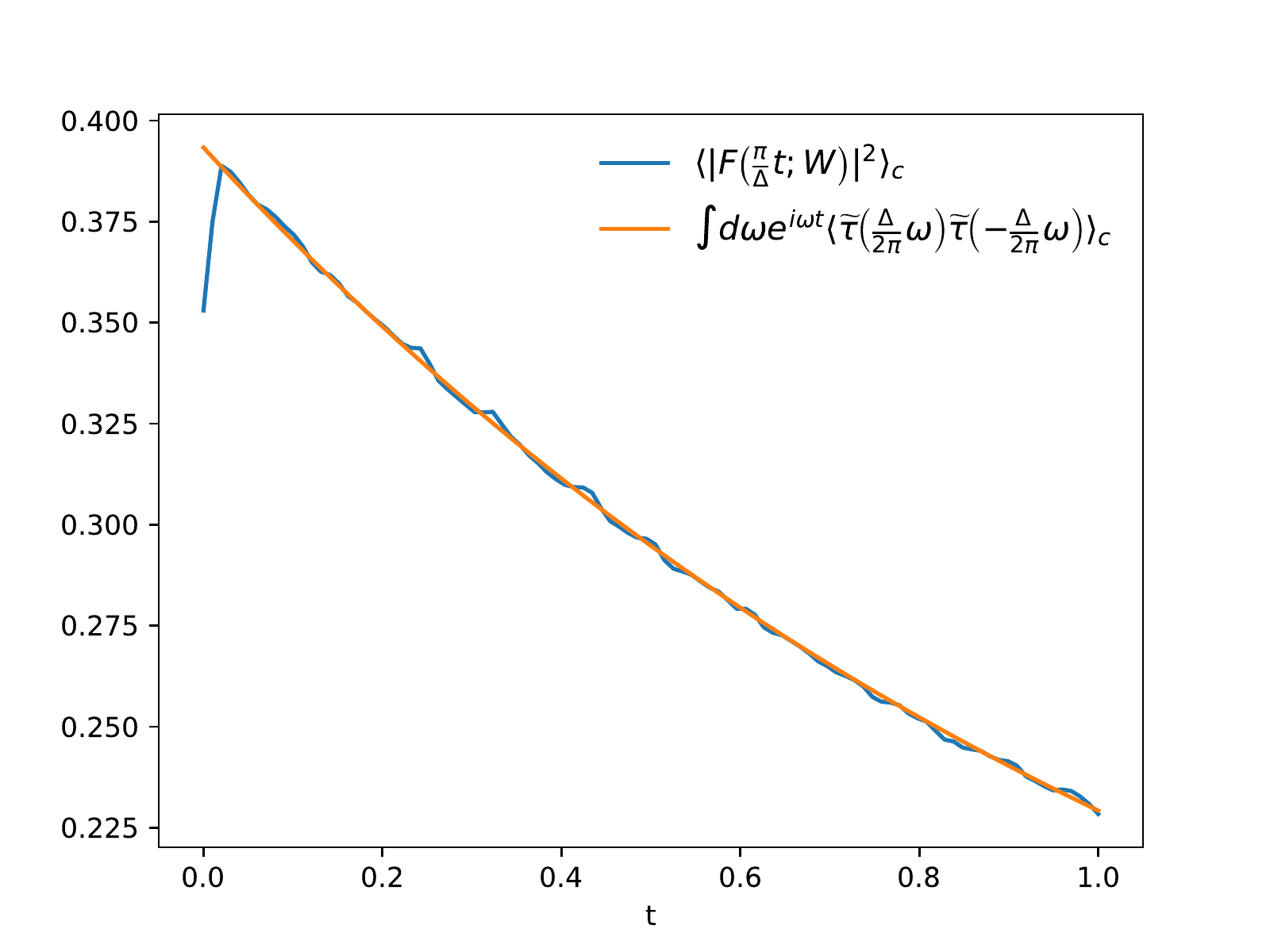}
        \caption{\small $M=1,L=1,g=g_{0}=1.25$}
        \label{fig:rtd_corr2}
    \end{subfigure}
    \caption{\small $C_{M,L}(t)$ of \ref{corrRTD1} compared with simulations of $300\times300$ random matrices, using \ref{eq:rtd_corr2} as an estimator. The small time discrepancy in the second figure appears to be a finite-size effect which occurs over a smaller window in time as the dimension of the matrices increases.}
    \label{fig:rtd_corr}
\end{figure}

The advantage of this approach is that the ensemble average of $F(t;W)$ converges faster than that of the RTD. Figures \ref{fig:rtd_corr1} and \ref{fig:rtd_corr2} compare the correlator obtained in this way from RMT simulations ($N=300$) with the prediction for $M=1$ and $L=0,1$.

\section{On extracting S-matrix zeros from scattering data} Let us now discuss a possibility of determining the positions of the zeros of a subunitary scattering matrix from an experiment, real or numerical. One may imagine two possibilities: either one has access to the total unitary $\mathcal{S}$ matrix, or only to its subunitary
observable/scattering block $S(E)$. Indeed, one may consider constructing a CPA in the form of a three port microwave network where one port plays the role of the attenuator providing absorption. In such a setup, if one disregards imperfections in the setup, the output in all three ports is directly accessible, hence the total $\mathcal{S}$ matrix.

In the usual case of $S-$matrix poles, one of the most well-established methods for determining their positions in the lower half-plane is the ``harmonic inversion'' \cite{wiersig_fractal_2008}, which estimates the poles by solving a set of non-linear equations. In a nutshell, one estimates a signal represented in the time domain as a  sum of decaying exponentials evaluated at times $t=n\tau$, where $\tau$ is the sample rate, by relying on the Pad\'{e} approximation:
\begin{align}
    c_{n} &:= \sum_{k=1}^{N}d_{k}e^{-iz_{k}n\tau},\\
    f(z) &:= \sum_{n=0}^{\infty}c_{n}z^{-n} = \sum_{k=1}^{N}\frac{d_{k}z_{k}}{z-z_{k}}\\
    &= \frac{P_{K}(z)}{Q_{K}(z)},
\end{align}
where $P_{K}/Q_{K}$ is the order $(K,K)$ Pad\'{e} approximant to $f(z)$. The zeros $z_{k}$ are the poles of $Q(z)$ and the amplitudes $d_{k}$ are the ratio of the residues and the poles. The integer $K$ is chosen as an upper bound to the number of true zeros $N$, leading to $N-K$ spurious zeros which must be discarded. In general, it is common to introduce a cut-off and discard zeros for which the magnitude of the residue falls below the cut-off. The procedure for  $S-$ matrix poles outlined in \cite{kuhl_resonance_2008} used both a cut-off and discarding those poles whose imaginary parts were smaller than the energy spacing (distance between energy samples), as well as  with real parts near the boundary of the sampling window.

Harmonic inversion can be performed directly on the $S-$matrix elements or, alternatively, on the Wigner time delay. When estimating zeros, the signal to be considered is given by the inverse of the determinant:

\begin{align}
   \frac{1}{\det S(E)} &=\prod_{n=1}^{N}\frac{E-E_{n}+i\Gamma_{n}/2}{E-\Re z_{n}-i\Im z_{n}}.\label{eq:signal1}
\end{align}

Alternatively, when the whole unitary $\mathcal{S}$ can be measured, one can use the expression \ref{eq:rtd} for the RTD. Assuming additionally a weak uniform absorption $\epsilon\ll1$ inside the scattering domain, the RTD can be alternatively computed from the unitary deficit of the determinant ratio, see \cite{fyodorov_reflection_2019} for a discussion:

\begin{align}
    \delta\mathcal{T}(E) &= -\frac{1}{\epsilon}\Re\log\frac{\det S(E+i\epsilon)}{\det S'(E+i\epsilon)}+O(\epsilon^{2}).\label{eq:signal2}
\end{align}

The advantage of using the RTD for extracting positions of complex zeroes is that the Lorentzians in the sum are all normalised to unity, providing us with a way to distinguish true and spurious zeros by looking at the corresponding residues. The residues in \ref{eq:signal1} involve instead a product over the remaining zeros and poles which can take arbitrary values. It is not a priori clear how to choose an appropriate cut-off, particularly when the zeros/poles appear as Lorentzians with varying amplitudes. We compare the performance of both methods for extracting the zeros by simulating $H_{0}$ from the GUE; parameters affecting the accuracy of the estimates are the number of samples $nE$ of $S(E)$ and the strength of the uniform absorption $\epsilon$. Since the first step of the harmonic inversion procedure is to take the Fourier transform, when using \ref{eq:signal1} we perform a second transform on the complex conjugate so that the zeros in both half-planes are accounted for. The detection of spurious zeros differs between the two methods. In the first method, we follow \cite{kuhl_resonance_2008} in using a cut-off and removing zeros near the boundaries. The cut-off has been chosen as 1, the value accounting well for most of the spurious zeros. In the second method, the zeros occur in complex conjugate pairs whose residues (after normalising by the energy spacing) should add up to 1.  We therefore grouped the zeros into conjugate pairs and removed those whose residues were significantly different from 1. This allowed us to bypass the need for the removal of zeros near the boundaries. Figure \ref{fig:hinv} shows the true zeros plotted against those estimated from \ref{eq:signal1} and \ref{eq:signal2}, for various parameter configurations. In figures \ref{fig:hinv1} and \ref{fig:hinv3}, there are what appear to be spurious zeros (isolated green crosses) among those estimated by the second method. These arise because of the appearance of two complex conjugate pairs for the same zero; in general we observe that as the strength of the uniform absorption $\epsilon$ increases, an increasing number of zeros are associated with two or more complex conjugate pairs. This is why these do not appear in figures \ref{fig:hinv2} and \ref{fig:hinv4}, where $\epsilon=10^{-6}$. One could attempt to group all pairs in order to bring the sum of resiudes closer to one, or simply discard all but one pair. Note also that in all four figures there are zeros near the boundaries which are discarded in the first method but included in the second.

\begin{figure}[ht]
    \centering
    \begin{subfigure}[t]{\columnwidth}
        \includegraphics[width=\linewidth]{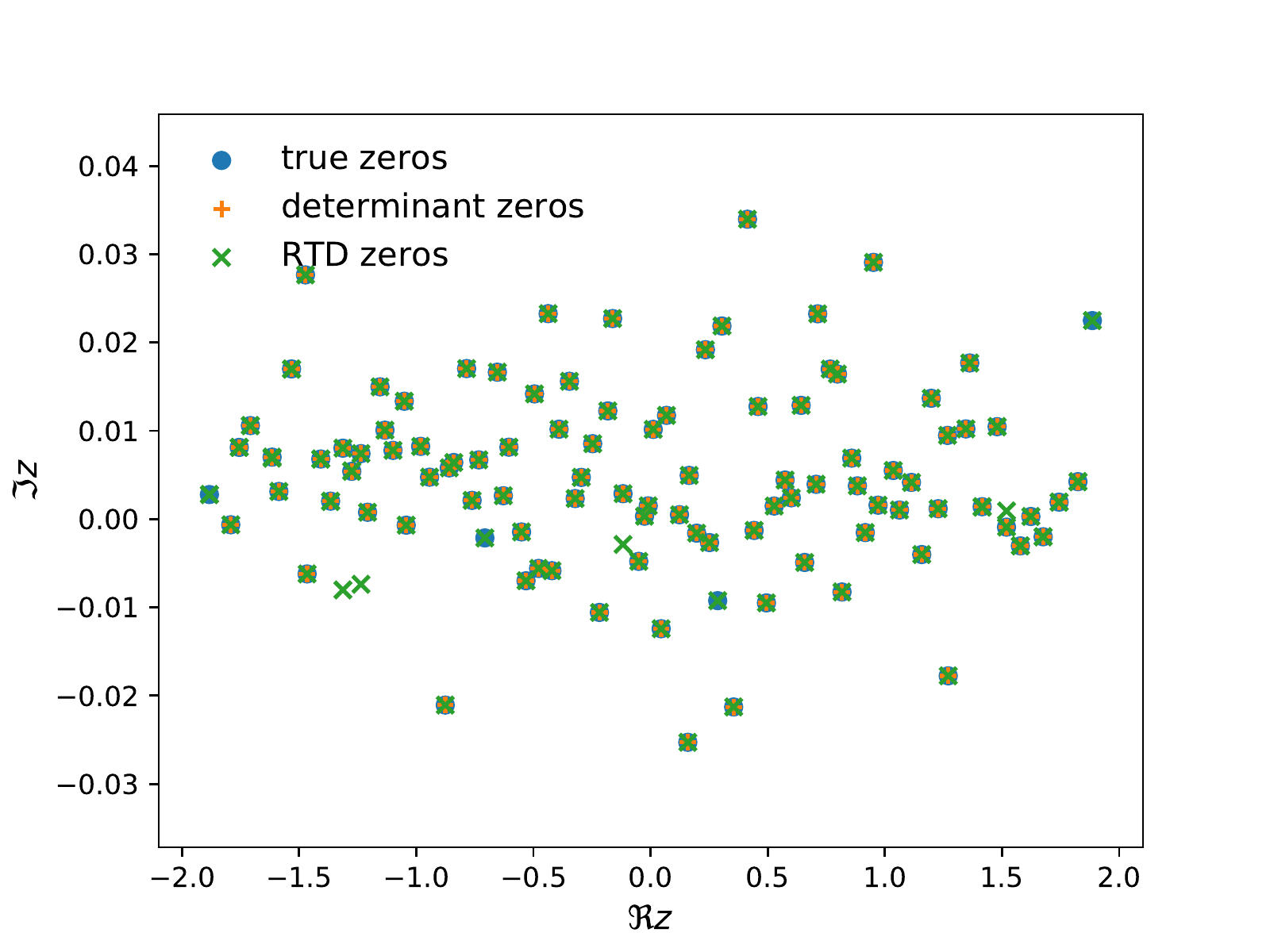}
        \caption{\small $\gamma=0.5,\rho=0.5,nE=800,\epsilon=10^{-5}$}
        \label{fig:hinv1}
    \end{subfigure}
    \begin{subfigure}[t]{\columnwidth}
        \includegraphics[width=\linewidth]{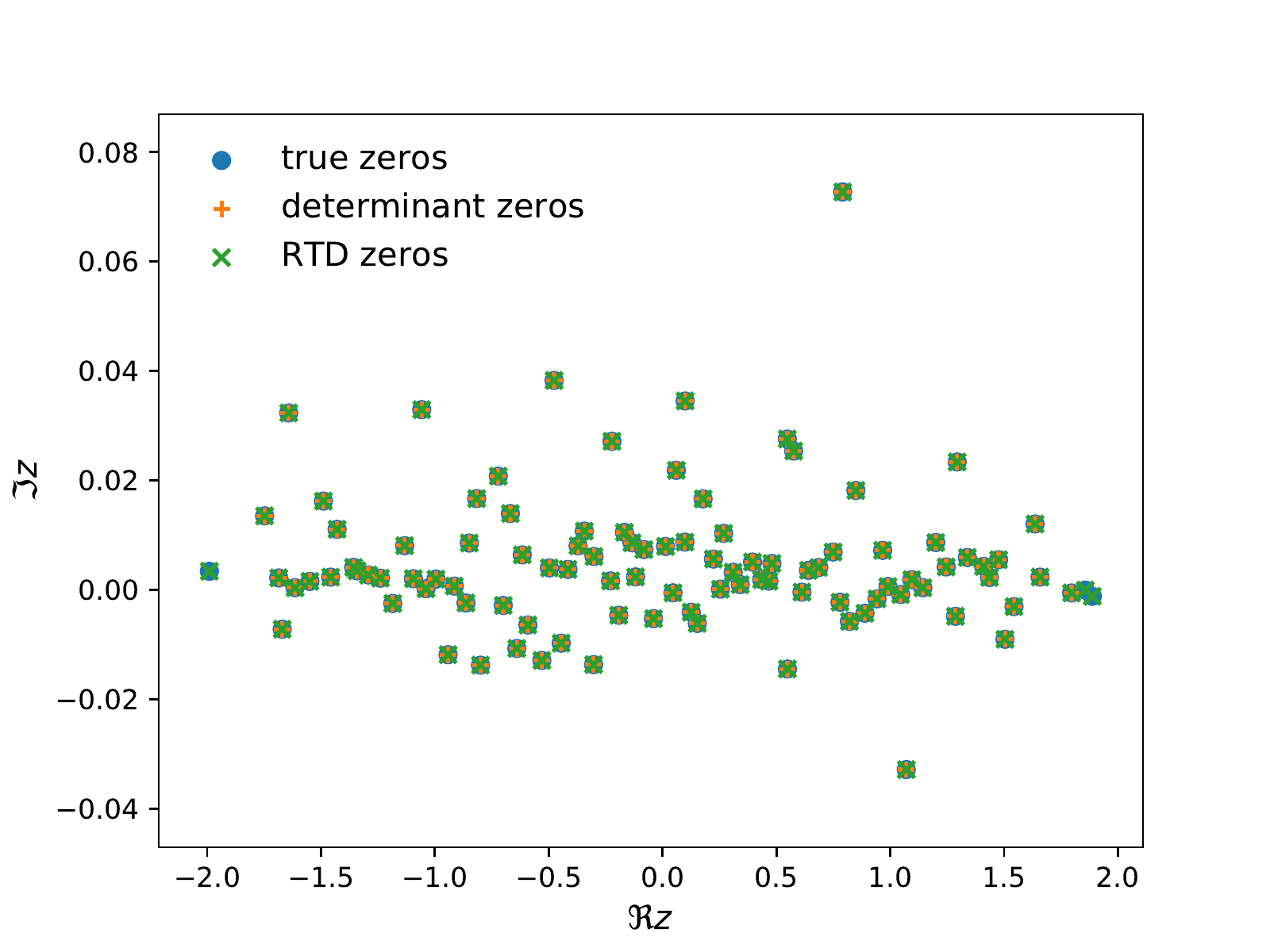}
        \caption{\small $\gamma=0.5,\rho=0.5,nE=800,\epsilon=10^{-6}$}
        \label{fig:hinv2}
    \end{subfigure}
\end{figure}
\begin{figure}[ht]
    \ContinuedFloat
    \begin{subfigure}[t]{\columnwidth}
        \includegraphics[width=\linewidth]{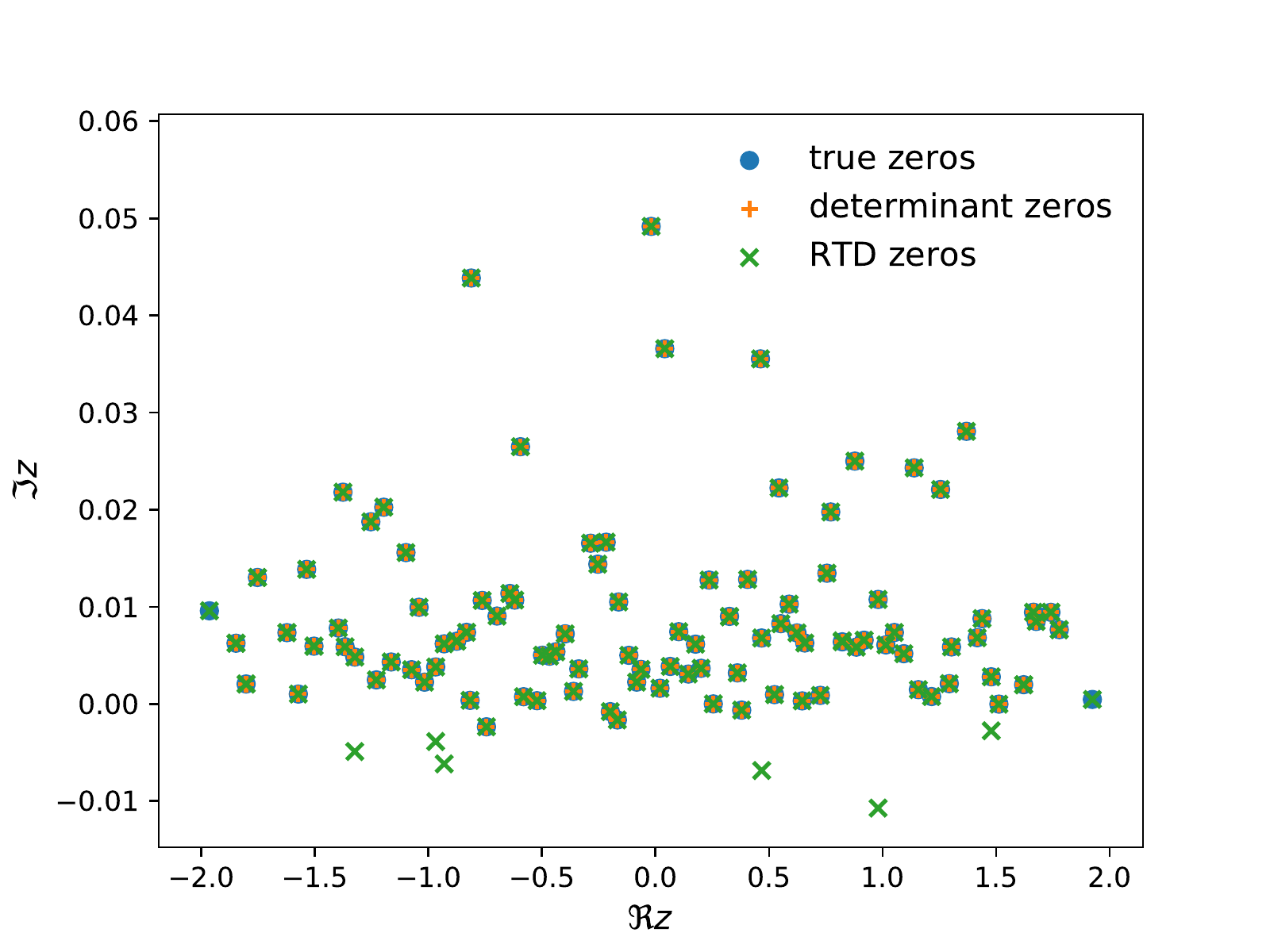}
        \caption{\small $\gamma=0.5,\rho=0.1,nE=800,\epsilon=10^{-5}$}
        \label{fig:hinv3}
    \end{subfigure}
    \begin{subfigure}[t]{\columnwidth}
        \includegraphics[width=\linewidth]{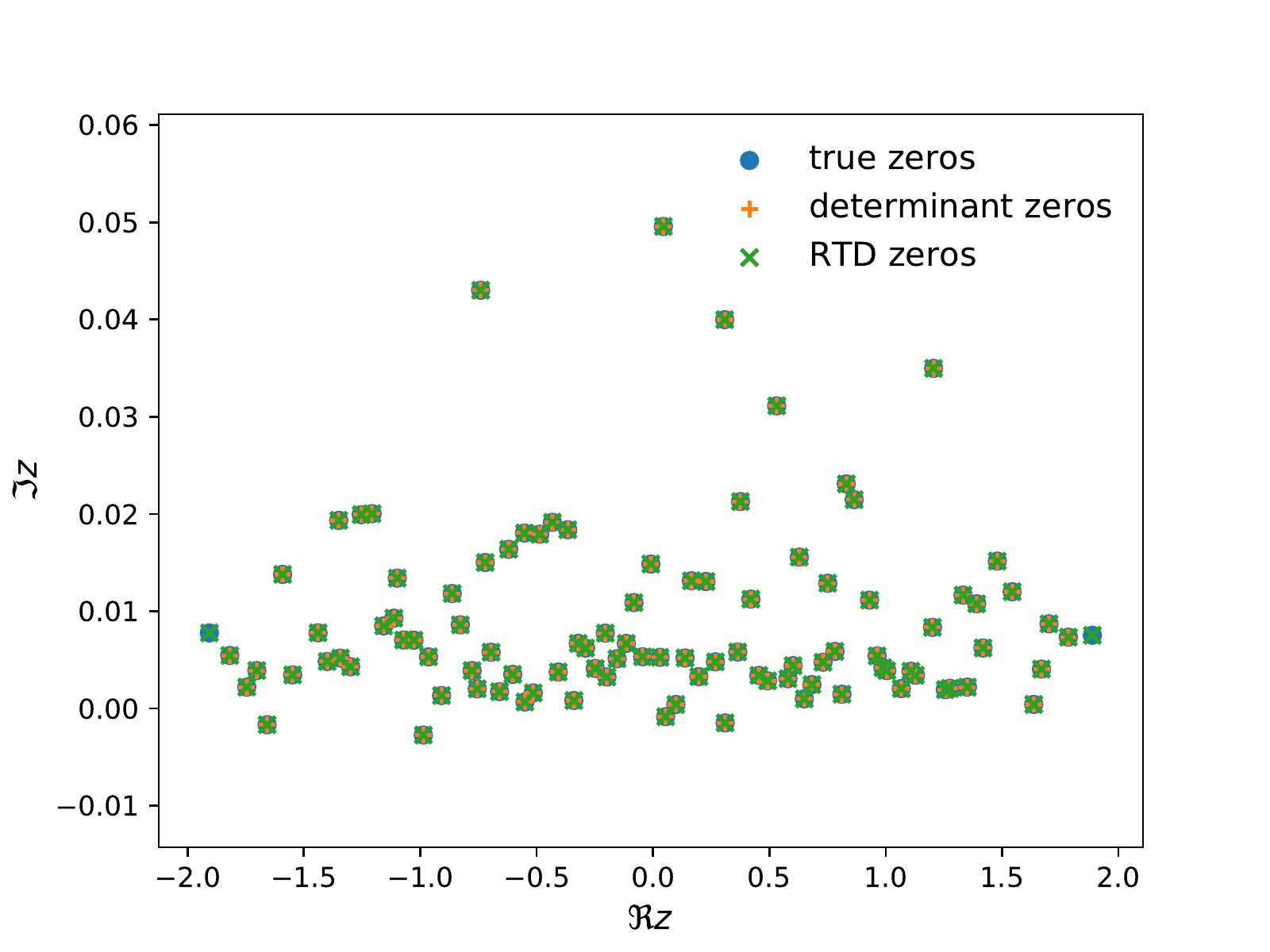}
        \caption{\small $\gamma=0.5,\rho=0.1,nE=800,\epsilon=10^{-6}$}
        \label{fig:hinv4}
    \end{subfigure}
    \caption{\small Estimated zeros using \ref{eq:signal1} (orange plus signs) and using \ref{eq:signal2} (green crosses) compared to true zeros (blue circles). In all cases, $N=100,M=2,L=1$.}
    \label{fig:hinv}
\end{figure}

\section{Discussion and open problems.} In conclusion, we have studied the statistics of the zeros of the subunitary $S$ matrix for a RMT-based model of scattering in chaotic cavities with localized absorption and broken time-reversal symmetry. The $n$-point correlation functions of the zeroes in the complex plane for a system with a finite number of scattering and absorbing channels ($M$ and $L$, respectively) can be calculated by extending the method used in  \cite{fyodorov_systematic_1999} for characterizing the $S-$matrix poles (formally equivalent to the $L=0$ case). The resulting kernel, in the limit of strong non-Hermiticity with at least one of the parameters $M$ and/or $L$ going to infinity, takes the form of a generalised Ginibre kernel, expected to be universal in this regime \cite{fyodorov_random_2003}. For finite $M<\infty$ and $L<\infty$, the kernel was used to obtain the Fourier transform of the correlation function $C_{M,L}(t)$ of the RTD, which has been shown to decay as
$C_{M,L}(t)\sim t^{-1}$ at large times for any $L>0$. This is in sharp contrast with the case $L=0$ (formally equivalent to replacing RTD with the Wigner time delay),  where a similar tail is known to be $M$-dependendent: $\sim t^{-M}$. In particular, this implies that the variance of the RTD is infinite regardless of the value of $M$ or $L$. We interpret this divergence as an evidence for the existence of a far tail $P(\widetilde{\delta\mathcal{T}})\sim\delta\mathcal{T}(E)^{-3}$
in the RTD probability density, and verify this in the regime of weakly open and weakly absorbing system.
We expect such a tail to be a superuniversal feature of the RTD distribition valid for all symmetry classes.

 The short time behaviour of the RTD correlation function  is again markedly different for $L=0$ and $L>0$: $C_{M,0}(t)\sim\frac{1}{4}t$ in the former case and $C_{M,L}(t)\sim C_{M,L}(0)+\dot{C}_{M,L}(0)t$ in the latter, with $C_{M,L}(0)\neq0$ and the coefficient $\dot{C}_{M,L}(0)$ depending on $g$ and $g_{0}$ in such a way that it changes from a positive value through zero to negative values as $g_{0}$ is reduced from infinity towards unity. This implies that there must exist a particular value of absorption parameter $g_{0}$ (depending on $g$) such that the large frequency asymptotic of the Fourier-transformed correlation function decays as $\omega^{-4}$ instead of the typical decay $\omega^{-2}$.

  We have also examined two methods for extracting the positions of $S-$matrix zeros from experimental scattering data by harmonic inversion of either the inverse determinant of $S$ or from the RTD. The first method is applicable when only the scattering part of $S$ is available, while for the second method the total $\mathcal{S}$ matrix must be accessible. The advantage of the second method is that we know in advance the residue of each zero, which allows us to distinguish more easily between true and spurious zeros. In the recent papers demonstrating the construction of a disordered CPA \cite{pichler_random_2019,chen_perfect_2020} the $S$ matrix was measured at regular intervals in an energy window, but rather than determine the zeros, each $S$ matrix was diagonalised and the eigenvalue closest to zero selected as a candidate for the CPA state. Using harmonic inversion one can instead directly estimate the zeros themselves.

Finally, let us mention that all non-perturbative treatment in our paper has been restricted to systems with broken time-reversal invariance. Actually, the mean density of complex eigenvalues $z_n$ for chaotic systems with preserved time-reversal invariance, with $H_0$ being taken from the Gaussian Orthogonal Ensemble, can be deduced for $L>0$ from known $L=0$ results by an ad hoc analytic contunuation, see \cite{fyodorov_distribution_2017}.
However its systematic controllable derivation, not speaking about deducing  the form of the two-point (and higher) correlation functions, remains an outstanding and challenging problem. This currently prevents us from any non-perturbative insights into statistics of the Reflection Time Difference (apart from its expected superuniversal
cubic tail in the probability density) for this important case.


\acknowledgments

{\bf Acknowledgments:} We thank Dr. Mikhail Poplavskyi for sharing his notes about the derivation of
(\ref{probdenzero}) and Prof. Steven Anlage for his interest in the project and valuable discussions of this and related topics. MO acknowledges funding from the EPSRC Centre for Doctoral Training in Cross-Disciplinary Approaches to Non-Equilibrium Systems (CANES) under grant EP/L015854/1.

\appendix

\section*{Appendix}

\subsection*{Derivation of the correlation function of $S-$matrix zeros.} We give a brief sketch of the derivation of the exprression \ref{eq:kernel} for the $n$-point correlation function. Since the derivation follows mainly  the steps of \cite{fyodorov_systematic_1999} (explained in more detail in \cite{fyodorov_random_2003}), we omit the detail and point out only the necessary modification. The object of study is the spectrum of the perturbed GUE matrix $J=H+i\Gamma$, where $\Gamma=\text{diag}(\gamma_{1},...,\gamma_{M+L},0,...,0)$ is a rank $M+L$ matrix where $\gamma_{1}\geq\cdots\geq\gamma_{M}>0>\gamma_{M+1}\geq\cdots\gamma_{M+L}$. It is the fact that the matrix
$\Gamma$ has eigenvalues of both signs which makes a difference from  \cite{fyodorov_systematic_1999}
and should be properly accounted for. The joint probability density for the matrix $J$ induced by $H$ is:

\begin{align}
    P(J)dJ &\propto e^{-\frac{N}{2}\text{tr}\left(\frac{J+J^{\dagger}}{2}\right)^{2}}\delta\left(\frac{J-J^{\dagger}}{2i}-\Gamma\right),
\end{align}

where $\delta(A)=\prod_{i,j}\delta(A_{ij})$. Making use of the Schur decomposition $J=U(Z+R)U^{\dagger}$, the integral over the upper triangular matrix $R$ can be performed with the delta function in the off-diagonal elements. The remaining delta functions in the diagonal elements are represented by a Fourier integral over a diagonal matrix $K$, leaving a Harish-Chandra/Itzykson/Zuber (HCIZ) integral with $K$ and $\Gamma$. Since $\Gamma$ is not of full rank, the limit of $N-M-L$ eigenvalues going to zero is calculated by repeated application of l'H\^{o}pital's rule to the original HCIZ formula. The result is the following expression for the density of eigenvalues $Z$:

\begin{equation}
    P(Z)  \propto \frac{\lvert\Delta(Z)\rvert^{2}}{\text{det}^{N-M-L}(\gamma)\Delta(\gamma)}
    e^{-\frac{N}{2}\Re\text{tr}Z^{2}-\frac{N}{2}\text{tr}\gamma^{2}}
\end{equation}
    \[
    \times  \int\frac{dK}{(2\pi)^{N}}\frac{e^{i\text{tr}K\Im Z}}{\Delta(K)}\,{\cal D}(K)
    \]
where we denoted  $\gamma=\text{diag}(\gamma_{1},...,\gamma_{M+L})$ and introduced
\begin{equation}
   {\cal D}(K)  :=
\end{equation}
\[
    \text{det}\begin{vmatrix}
    e^{-ik_{1}\gamma_{1}}&\cdots&e^{-ik_{1}\gamma_{M+L}}&(-ik_{1})^{N-M-L-1}&\cdots&1\\
    \vdots&\ddots&\vdots&\vdots&\ddots&\vdots\\
    e^{-ik_{N}\gamma_{1}}&\cdots&e^{-ik_{N}\gamma_{M+L}}&(-ik_{N})^{N-M-L-1}&\cdots&1
    \end{vmatrix}.
\]
 The difference now from the original derivation is that the terms $e^{-ik_{j}\gamma_{c}}$ are represented by the integral:
\begin{align}
    e^{-ik_{j}\gamma_{c}} &= \frac{1}{2\pi i}\int_{\mathcal{L}_{c}}\frac{e^{-ik_{j}\lambda_{c}}}{\lambda_{c}-k_{j}}d\lambda_{c},
\end{align}
where $\mathcal{L}_{c}=\text{sign}(\gamma_{c})(-\mathbb{R}+i0)$. After taking the $\lambda$ integrals outside the determinant and using the following identity:
\[
    \det\begin{vmatrix}
    \frac{1}{\lambda_{1}-k_{1}}&\cdots&\frac{1}{\lambda_{\mathcal{M}}-k_{1}}&(-ik_{1})^{N-\mathcal{M}-1}&\cdots&1\\
    \vdots&\ddots&\vdots&\ddots&\vdots\\
    \frac{1}{\lambda_{1}-k_{N}}&\cdots&\frac{1}{\lambda_{\mathcal{M}}-k_{N}}&(-ik_{N})^{N-\mathcal{M}-1}&\cdots&1
    \end{vmatrix}
 \]
 \begin{align}
     \propto \frac{\Delta(\Lambda)\Delta(K)}{\prod_{j=1}^{\mathcal{M}}\prod_{i=1}^{N}(\lambda_{j}-k_{i})},
\end{align}
which follows by elementary row and column operations, the final expression for the density $P(Z)$ is:
\begin{align}
    P(Z) &\propto \frac{1}{\det^{N-\mathcal{M}}(\gamma)\Delta(\gamma)}\lvert\Delta(Z)\rvert^{2}e^{-\frac{N}{2}\text{Re}\text{tr} Z^{2}-\frac{N}{2}\text{tr}\gamma^{2}}\nonumber\\\
    &\quad\times\int_{\mathcal{L}_{1}}\cdots\int_{\mathcal{L}_{\mathcal{M}}}d\Lambda\Delta(\Lambda)
    e^{-i\sum_{j=1}^{\mathcal{M}}\gamma_{j}\lambda_{j}} \nonumber
   \\ & \times \prod_{i=1}^{N}\sum_{j=1}^{\mathcal{M}}\frac{e^{i\lambda_{j}\text{Im} z_{i}}}{\prod_{l\neq j}(\lambda_{l}-\lambda_{j})}\theta(\text{Im}\lambda_{j}\text{Im} z_{i}),\label{probdenzero}
\end{align}

From this point onwards the derivation follows exactly along the lines of \cite{fyodorov_random_2003}, where in the appendix of that paper the $n$-point correlation function is related to the average of a product of characteristic polynomials which is subsequently evaluated by integrating over anti-commuting variables.

\subsection*{Distribution of RTD in the regime of weak coupling and absorption.} In this regime the imaginary part
of the zeroes is much smaller than the typical separation between the real parts, and, moreover,
imaginary parts of neighbouring zeroes are independent to the leading order.
Then the dominant contribution to RTD at a given real energy $E$ can be estimated by a heuristic argument \cite{fyodorov_statistics_1997} that takes into account only the zero whose real part is the closest to the energy value $E$ in the sum of Lorentzians \ref{eq:rtd}. Attributing the index $n$ to this particular zero and
 defining $u_{n}=\frac{\beta\pi}{\Delta}(E-\Re z_{n})$ and $y_{n}=\frac{\beta\pi}{\Delta}\Im z_{n}$, we have

\begin{align}
    \widetilde{\delta\mathcal{T}}(E) &\simeq \frac{2y_{n}}{u_{n}^{2}+y_{n}^{2}}.
\end{align}

The real and imaginary parts of $z_{n}$ are independent in the weak coupling regime, with the latter having the density:

\begin{align}
    P^{\beta}_{Y}(y) &= \int \frac{dk}{2\pi}\frac{e^{iky}}{\prod_{a=1}^{M}(1+2ik\gamma_{a})^{\beta/2}\prod_{b=1}^{L}(1-2ik\rho_{b})^{\beta/2}}.
\end{align}

We also assume that $u_{n}$ is uniformly distributed in $[-\beta\pi,\beta\pi]$. The result of these approximations is the following estimate for the density of $\widetilde{\delta\mathcal{T}}\equiv\tau$:

\begin{align}
    P_{\widetilde{\delta\mathcal{T}}}^{\beta}(\tau) &= \int dy P_{\beta}(y)\int_{-\beta\pi}^{\beta\pi}du\delta\left(\tau-\frac{2y}{u^{2}+y^{2}}\right)\\
    &\simeq\frac{2}{\tau^{3}}\int_{0}^{\min(1,\frac{\beta^{2}\pi^{2}\tau^{2}}{4})}\sqrt{\frac{y}{1-y}}
    \left[\theta(\tau)P^{\beta}_{Y}\left(\frac{2y}{\tau}\right) \right.
\end{align}
\[
\left.+\theta(-\tau)P_{Y}^{\beta}\left(-\frac{2y}{\tau}\right)\right]dy.
\]

Focusing on positive $\widetilde{\delta\mathcal{T}}$
 and equivalent channels $\gamma_{a}=\gamma\ll 1,\rho_{b}=\rho\ll 1$, let us first assume that 
 $\rho\gtrsim \gamma$.  Then  
 we can identify two regimes for $\tau>2/(\beta\pi)$: i) $\tau\gamma\ll1$ and ii) $\tau\gamma\gg1$. In the first regime, the dominant term in $P^{\beta}_{Y}(2y/(M\tau))$ is $y^{\beta M/2-1}e^{-\frac{y}{\tau\gamma}}$:

\begin{align}
    P^{\beta}_{\widetilde{\delta\mathcal{T}}}(\tau) &\simeq\frac{\gamma^{\beta L/2-1}}{\tau^{3}\Gamma(\beta M/2)(\gamma+\rho)^{\beta L/2}}
    \\ \nonumber & \times
    \int_{0}^{\infty}\sqrt{y}\left(\frac{y}{\tau\gamma}\right)^{\beta M/2-1}e^{-\frac{y}{\tau\gamma}}dy\\
    &=\frac{\gamma^{\beta(L+1)/2}}{(\gamma+\rho)^{\beta L/2}}\tau^{-3/2},
\end{align}

where we have set the upper limit of integration to infinity and taken $1-y\simeq1$, both justified by the exponential damping. In the second regime, the dominant term is now just $e^{-\frac{y}{\tau\gamma}}$, which is approximately unity:

\begin{align}
    P^{\beta}_{\widetilde{\delta\mathcal{T}}}(\tau) &\simeq \frac{\Gamma(\beta(M+L)/2-1)}{\Gamma(\beta M/2)\Gamma(\beta L/2)}\frac{1}{\tau^{3}\gamma^{\beta M/2}\rho^{\beta L/2}}
 \\ \nonumber & \times
    \left(\frac{\gamma\rho}{\gamma+\rho}\right)^{\beta(M+L)/2-1}\int_{0}^{1}\sqrt{\frac{y}{1-y}}\\
    &=\frac{\pi\Gamma(\beta(M+L)/2-1)}{2\Gamma(\beta M/2)\Gamma(\beta L/2)}
   \\\nonumber & \times \frac{1}{\gamma^{\beta M/2}\rho^{\beta L/2}}\left(\frac{\gamma\rho}{\gamma+\rho}\right)^{\beta(M+L)/2-1}\tau^{-3}.
\end{align}
 Thus we see that $P^{\beta}_{\widetilde{\delta\mathcal{T}}}(\tau)$ behaves as a power law with universal exponents,
summarized in (\ref{unitail}). 

If one however assumes $\rho\ll \gamma$ and repeats the analysis above, one finds that the regime
 $\tau\gg 1/\gamma$ should be further subdivided into two new regimes: 
$ 1/\gamma \tau\ll 1/\rho$ and $\tau\gg 1/\rho$. Only in the latter case one reproduces the superuniversal asymptotics $P^{\beta}_{\widetilde{\delta\mathcal{T}}}(\tau) \simeq \tau^{-3}$ whereas in the former the asymptotic
is changed to  $P^{\beta}_{\widetilde{\delta\mathcal{T}}}(\tau) \simeq \tau^{-(2+\beta M/2)}$. 
This type of tail behaviour is precisely one which is typical for the Wigner time delays, to which 
the RTD formally reduces as $\rho\to 0$.   

The same arguments can be made for negative $\widetilde{\delta\mathcal{T}}$ by exchanging the role of $\gamma$ with that of $\rho$.

\end{document}